# Immunity to *Plasmodium knowlesi* H strain malaria in olive baboons

Barasa Mustafa[1,2*], Maamun Jenneby[2], Kagasi Ambogo Esther[2],
Ozwara Suba Hastings[2], Gicheru Muita Michael[1]

[1]*Zoological Sciences Department, School of Pure and Applied Sciences, Kenyatta University, Nairobi*
*Department of Tropical and Infectious Diseases, Institute of Primate Research, Nairobi, Kenya*

**ABSTRACT**

Malaria disease is a major global health and economic development threat. It results in approximately 2.7 million deaths annually. There is currently no vaccine that has been licensed for use against malaria. Studies in animal models, especially non-human primates can lead to the revelation of possible immunological mechanisms that can lead to protection or predisposition of the host to malaria. *Plasmodium knowlesi,* a simian and human malaria parasite, is an attractive experimental parasite for malaria research since it can infect olive baboons (*Papio anubis*), non-human primates that have similar host-pathogen interactions to humans. This study was carried out to determine host immunological profiles provoked in olive baboons during the course of an infection with *Plasmodium knowlesi*. A total of eight adult baboons were intravenously inoculated with overnight cultured blood stage *P. knowlesi* H strain parasites. Five of these baboons became acutely infected while the other three became chronically infected. The immunological basis of this dual outcome of the infection was determined by measuring circulating cytokine (T helper 1 and T helper 2) and antibody (immunoglobulin G and immunoglobulin M) responses elicited in the infected baboons on a weekly basis by Enzyme Linked Immunosorbent Assay (ELISA) for up to six weeks post infection. Generated data for the first time indicated that acute *P. knowlesi* malaria is accompanied by increased concentrations of interferon gamma (IFN γ), tumour necrosis factor alpha (TNF α) and IL 6 and reduced levels of circulating interleukin 10 (IL 10), IL 4, IL 12, immunoglobulin G (IgG) and IgM in the baboon host. These results are largely agreeable with data from human studies, thereby increasing the relevance of the olive baboon - *P. knowlesi* experimental infection system for future malaria studies.

**Key words**: Baboon, *Plasmodium knowlesi*, malaria, immunoglobulin, cytokine, immunity.

## INTRODUCTION

The evaluation of new drugs and vaccines for use in humans requires testing in animal models that develop a disease comparable to that in humans. The baboon is an attractive experimental model because its well characterised and is also frequently used in biomedical research. Baboons are fully susceptible to infection by *P. knowlesi*, a parasite that shares many of the vaccine candidate molecules with *P. vivax* and whose entire genome has been sequenced by up to over 5-fold coverage (Ozwara *et al*., 2003). The disease profiles reported in baboons are similar to those in rhesus monkeys following experimental infection with *P. knowlesi* (Ozwara *et al*., 2003), showing that the virulence of the strain is similar in both animal models. The urgency to determine non-human primate immune responses to *P. knowlesi* is further increased by the fact that incidences of human *P. knowlesi* infections are on the rise (Cox-Singh *et al.,* 2008; Ng *et al.,* 2008). When infected with *P. knowlesi,* the *Papio anubis* species of baboons show various clinical aspects that are also seen in human malaria, including cerebral involvement (Aikawa *et al.,* 1990; Ozwara *et al*., 2003). *Plasmodium knowlesi* has been instrumental in the discovery and characterization of malaria blood stage vaccine candidate AMA-1 (Kocken *et al*., 1999). It has also been used to identify, develop, and evaluate vaccine and drug candidates (Kocken *et al*., 1999). Recently, protocols for genetic modification of *P. knowlesi* have been developed. These are powerful tools for drug and vaccine development because they enable the functions of target drug and vaccine candidates to be screened using *P. knowlesi* genes (Ozwara *et al*., 2003). The use of the baboon is a pivotal undertaking in expanding the number of primate hosts for use in *in vivo* experiments for studying host-parasite interactions of both the wild type and the transfected *P. knowlesi* parasites. Humans and baboons have similar immunopathophysiology and host-pathogen interactions and therefore data acquired from baboon experiments could provide information on what could happen in a human situation (Ozwara *et al*., 2003). Before this report, cytokine, immunoglobulin G and immunoglobulim M immune responses elicited in baboons (*Papio anubis*) against *P. knowlesi* had not been described yet the determination of the immunity (humoral and cytokine immunity) of the baboons to *P. knowlesi* infections is necessary towards its development as a model for the analysis of malarial host-parasite interactions. This current study investigated for the first time, some of the immunological responses (antibody, T helper 1 and T helper 2 cytokines) to experimental *P. knowlesi* H strain infections in olive baboons.

*Corresponding author:
Mustafa Barasa, MSc.
Zoological sciences Department,
School of Pure and Applied Sciences,
P. O. Box 43844, Kenyatta University, Nairobi, Kenya
Email: mustrech@yahoo.com

## MATERIALS AND METHODS

### Experimental baboons and parasites

Experimental animals consisted of 8 adult baboons (*Papio anubis*) originally from Kajiado District of Kenya with a mean weight of 10 Kg. These baboons were trapped and maintained in the quarantine facilities at the Institute of Primate Research (IPR). All experiments were performed in accordance with the institutional Scientific and Ethical Review Committee (ISERC) guidelines of IPR. The baboons were examined and confirmed free of hemoprotozoan, gastrointestinal parasites and Simian Immunodeficiency Virus (SIV) before being included in the study. Additionally microbiological examination of effusions, pus, ulcer material and skin specimens was done to detect pathogenic agents which cause infection in wounds and the skin. Overnight cultured *P. knowlesi* H strain parasites were used to initiate blood stage malaria infections in experimental baboons in a bio-containment facility. The complete parasite culture medium consisted of 2.5% baboon erythrocyte PCV, 20% baboon serum, 15 μg/ml gentamycin solution and the rest RPMI 1640 (Sigma-Aldrich, USA). All baboon sera for use had been heat inactivated previously at 56°C. One hundred to two hundred microlitres of culture were used for thin smear preparation (for starting and daily parasitaemia). Cultures were mixed gently and transferred into sterile T 25 culture flasks. For the gassing of cultures, new gassing needles were heat sterilized before connection to a gas pipe fitted with a 0.2 μm filter and the cultures were gassed for 25 seconds. The gases in the mixture were (5% $CO_2$, 5% $O_2$ and 90% $N_2$). The flasks were tightly capped and transferred to an incubator (37°C). For blood stage inoculations, the parasites were resuspended to a population of $1.0 \times 10^6$ parasites/ml in incomplete RPMI 1640 (GIBCO) and injected into the baboons through the femoral vein. Parasitaemia and clinical symptoms were monitored on a daily basis to determine disease profiles (Ozwara *et al.,* 2003). At 10 % level of parasitaemia baboons were intravenously injected with chloroquin sulphate at a dosage of 5 mg /kg daily for 3 days. The baboons were bled weekly for up to six weeks for the extraction of plasma and sera for the determination of antibody and cytokine responses. Baseline sampling (termed as week zero in graphical data) was also done for the quantification of the same immunological parameters. For all invasive procedures, the baboons were anaesthetised with ketamine hydrochloride (10mg/kg).

### Immunoglobulin G (IgG) and Immunoglobulin M (IgM) Measurement

Immunoglobulin G (IgG) and Immunoglobulin M (IgM) concentrations were measured using ELISA in order to determine the level of humoral immune responses provoked in the baboons. Ninety six well flat bottomed ELISA microtiter plates (Sigma-Aldrich, USA) were coated using 50 μl/well of carbonate bicarbonate buffer (pH 9.6) containing $10^8$ (parasites used in antigen preparation) crude sonicated *P. knowlesi* parasite antigen. These were incubated overnight at 4°C for adsorption. Excess coating buffer was then flicked of and 100 μl/well of blocking buffer (3% BSA in PBS) added using a multichannel pipette followed by incubation for 1 hour at 37°C. The plates were then washed six times using wash buffer (0.05% tween in PBS) delivered by an automatic washing machine and sera samples added in triplicate (1: 100 dilution in 50 μl/well) on ice before incubation for one hour as before. The plates were then washed as before and 50 μl/well of 1:2000 dilution of either antihuman Immunoglobulin G horseradish peroxidase or antihuman Immunoglobulin M horseradish peroxidase (HRP; Sigma, USA) added. One hr incubation at 37°C then followed after which the plates were washed ten times using washing buffer. Colour development was achieved by adding 50 μl/well of Tetramethylbenzidene (TMB) substrate (Sigma, USA) and optical densities were read using a Dynatech MRX ELISA reader at 630 nm filter setting after 15 min of incubation at 37°C (Gicheru *et al.,* 1995; Barasa *et al.,* 2010).

### Determination of cytokine responses in sandwich ELISA

Cytokine ELISA was done in order to measure the levels of cytokine mediated immune responses induced in the baboons. The assays were performed essentially as reported previously (Gicheru *et al.,* 1995; Barasa *et al.,* 2010b). Ninety six well flat bottomed ELISA microtiter plates (Sigma-Aldrich, USA) were coated with 5 μg/ml of cross reactive antihuman/baboon cytokine (IFN γ, TNF α IL 4, IL 6, IL 10, IL 12) capture monoclonal antibody (Becton Dickinson) delivered 50 μl/well. These were incubated overnight at $4^0$C. Excess coating buffer was then flicked off and the wells blocked with 100 μl/well blocking buffer (3% BSA in PBS) followed by 1 hr incubation at 37°C. After washing the plates six times using ELISA washing buffer (0.05% tween in PBS), undiluted sera samples and recombinant cytokine standards were dispensed in duplicate, 50 μl/well and plates incubated for 2 hr at 37°C.

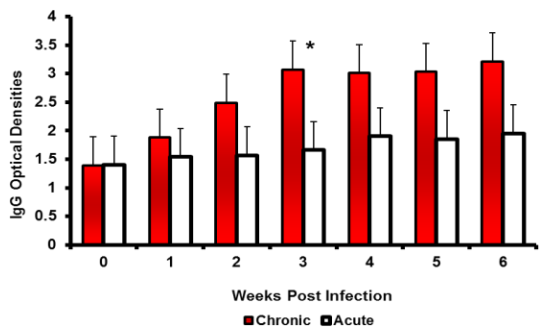

**Fig 1:** Anti *P. knowlesi* IgG levels in acute and chronic baboons infected with *P. Knowlesi*. *: Asterisk means significant difference between IgG levels in acute and chronic baboons, $P < 0.05$; n: acute = 5; chronic = 3.

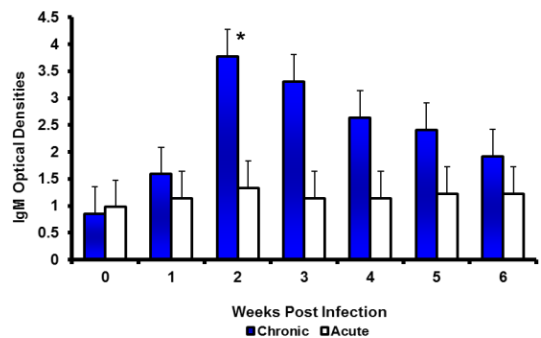

**Fig 2:** Anti *P. knowlesi* IgM levels in acute and chronic baboons infected with *P. Knowlesi*. *: Asterisk means significant difference between IgM in acute and chronic baboons, $P < 0.05$; n: acute = 5; Chronic = 3.

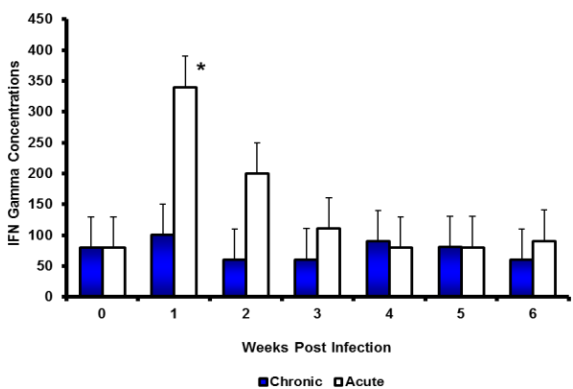

**Fig 3:** Interferon gamma responses in acute and chronic *P. knowlesi* infected baboons. *: Asterisk means significant difference in IFN gamma responses in chronic and acutely infected baboons, $P < 0.05$. n: acute = 5; Chronic = 3. Interferon gamma dominated the T helper 1 response in acute baboons.

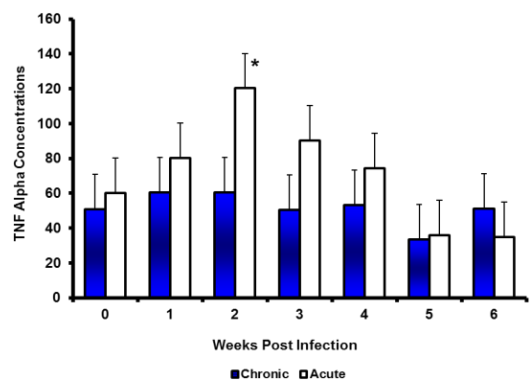

**Fig 4:** Tumour necrosis factor alpha responses in acute and chronic *P. knowlesi* infected baboons. *: Asterisk means significant difference in TNF alpha responses in acute and chronic baboons, $P < 0.05$; n: acute = 5; Chronic = 3.

Standards were serial diluted by transferring 50 µl from well to well with mixing beginning with a neat concentration of 500 pg/ml. The plates were then washed as before and detector mouse biotinylated antibaboon cytokine (IFN γ, TNF α IL 4, IL 6, IL 10, IL 12) monoclonal antibodies (Becton Dickinson) added 50 µl/well at dilutions of 1:2000. This was followed by a one hr incubation period at 37°C then washing as before. Streptavidin Horse Radish Peroxidase (HRP; Sigma, USA) diluted 1:2000 was added 50 µl/well and incubated 1 hr at 37°C followed by washing six times. Colour development was achieved by adding 50 µl/well of Tetramethylbenzidene (TMB) substrate (Sigma, USA) and optical densities were read using a Dynatech MRX ELISA reader at 630 nm filter setting after 15 min of incubation at 37 °C (Gicheru *et al.*, 1995).

### Statistical analysis

The infected baboons were grouped as either chronically infected or acutely infected depending on parasitaemia quantities experienced. Baboons that had peak parasitaemia measurements of below 5% were classified as chronically infected while baboons that had parasitaemia measurements above 5% were classified as acutely infected. Mean values of antibody and cytokine levels of the chronically infected group of baboons were compared with the respective mean values of the acutely infected group of baboons using non parametric Mann-Whitney U analysis. Probability values of *$P < 0.05$* were considered significant.

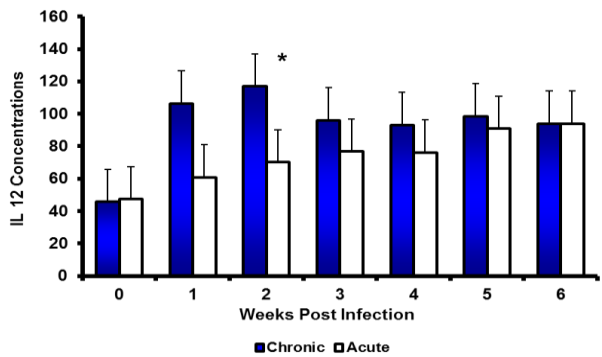

**Fig 5:** Interleukin 12 responses in acute and chronic *P. knowlesi* infected baboons. *: Asterisk means significant difference between IL 12 levels in acute and chronic baboons, $P < 0.05$; n: Acute = 5; Chronic = 3. Interleukin 12 dominated the T helper 1 response in the chronically infected baboons.

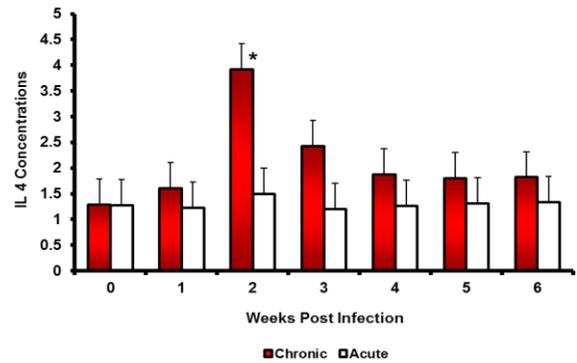

**Fig 6:** Interleukin 4 responses in acute and chronic *P. knowlesi* infected baboons. *: Asterisk means significant difference between acute and chronic baboons in their IL 4 responses, $P < 0.05$. *: Asterisk means significant difference between IL 4 in acute and chronic baboons, $P < 0.05$; n: Acute= 5; Chronic = 3.

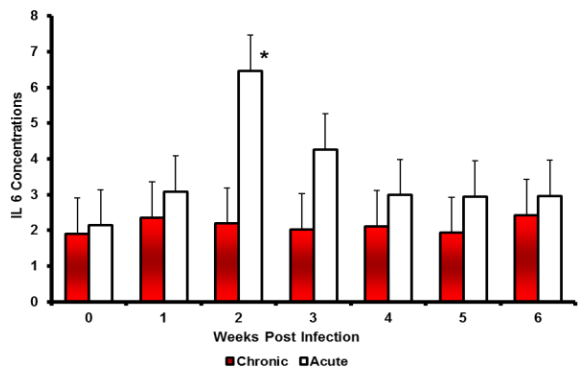

**Fig 7:** Interleukin 6 responses in acute and chronic *P. knowlesi* infected baboons. *: Asterisk means significant difference between IL 6 levels in acute and chronic baboons, $P < 0.05$; n: Acute = 5; Chronic = 3.

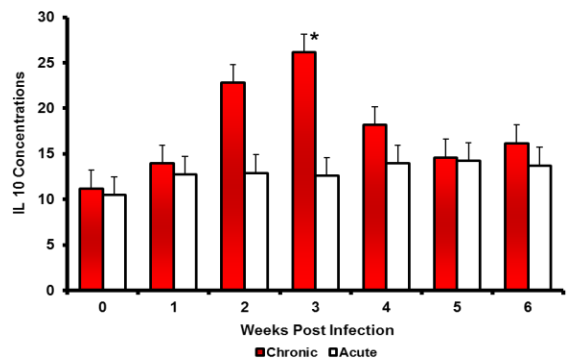

**Fig 8:** Interleukin 10 responses in acute and chronic *P. knowlesi* infected baboons. *: Asterisk means significant difference between IL 10 levels in acute and chronic baboons, $P < 0.05$; n: Acute = 5; Chronic = 3. Interleukin 10 was the most highly detected of the T helper 2 cytokines in both acute and chronic baboons.

## Parasitaemia development and occurrence of a dual outcome of infection

All the baboons inoculated with *P. knowlesi* H strain blood stages developed patent parasitaemia with all the blood stages being manifested. Five baboons became acutely infected while the rest (three) became chronically infected. Acute animals developed severe malaria symptoms and had high parasitaemia that systematically increased to over 5%, reaching as high as 10.3% in some cases at the time of treatment. Chronic animals had peak parasitaemia levels ranging from 2.3%-5%, which thereafter decreased to less than 2.2 % parasitaemia on average. Mean peak parasitaemia occurred at day 7 post infection in both chronically and acutely infected baboons. Parasitaemia differences were significant between the two groups ($P < 0.05$).

Baboons that developed acute infection displayed severe malaria symptoms that included apathy, ruffled hair, loss of appetite and vomiting while chronically infected baboons had mild asymptomatic malaria that manifestedin having low to moderate symptoms as observed in severely infected animals.

## Immunoglobulin G and Immunoglobulin M responses

Mean OD (Optical Density) levels of IgG in chronically infected baboon ranged from 1.39 to 3.21 while in acutely infected baboons they ranged from 1.45 to 1.95. Chronically infected baboons detectably produced significantly higher levels of IgG than acutely infected baboons (Fig 1; $P < 0.05$).

In chronically infected baboons IgM OD levels ranged from 0.85 to 3.77 while in acute ones levels ranged from 0.98 to 1.14. Chronically infected baboons therefore raised significantly higher levels of IgM than acutely infected baboons (Fig 2; $P < 0.05$).

## T helper 1 cytokine responses

In chronically infected baboons IFN γ levels ranged between 60.33 pg/ml and 100.56 pg/ml while in acutely infected baboons levels ranged between 80.11 pg/ml and 340.05 pg/ml. Acutely infected baboons produced significantly higher levels of IFN γ than chronically infected baboons (Fig 3; $P < 0.05$). Tumour necrosis factor alpha levels in chronically infected baboons ranged from 33.66 pg/ml to 60.55 pg/ml while in acute ones levels ranged from 34.87 pg/ml to 120.29 pg/ml. Acutely infected baboons manifested with significantly higher levels of TNF α than chronically infected baboons (Fig 4; $P < 0.05$). Interleukin 12 levels in chronically infected baboons ranged from 45.77 pg/ml to 116.87 pg/ml while in acute ones levels ranged from 47.44 pg/ml to 93.98 pg/ml. Chronically infected baboons had significantly higher levels of IL 12 than acute baboons (Fig 5; $P < 0.05$).

## T helper 2 cytokine responses

Interleukin 4 concentration in chronically infected baboons ranged from 1.29 pg/ml to 3.92 pg/ml while in acutely infected baboons it ranged from 1.2 pg/ml to 1.34 pg/ml. Therefore chronically infected baboons produced significantly higher levels of IL 4 than acutely infected baboons (Fig 6; $P < 0.05$). In chronically infected baboons IL 6 levels ranged from 1.90 pg/ml to 2.42 pg/ml while in acutely infected baboons the range was from 2.14 pg/ml to 6.47 pg/ml. There was a significantly higher level of IL 6 in acutely infected baboons than in chronically infected baboons (Fig 7; $P < 0.05$). In acutely infected baboons levels of IL 10 ranged from 10.48 pg/ml to 14.21 pg/ml while in chronically infected baboons levels ranged from 11.20 pg/ml to 26.16 pg/ml. Chronically infected baboons therefore expressed significantly higher levels of IL 10 than acutely infected baboons (Fig 8; $P < 0.05$).

## DISCUSSION

This study was designed to determine the immunological profiles (antibody and cytokine profiles) induced in the baboon (*Papio anubis*) host following experimental infection by *Plasmodium knowlesi* H strain parasites. Some baboons became acutely infected while others became chronically infected (Ozwara *et al.,* 2003). For a majority of the immunological parameters analysed, significant differences were detected between chronic and acute malaria baboons over this infection timeline. Immunoglobulin G and immunoglobulin M concentrations were measured to determine their roles in determining disease/infection profiles during such infections. Chronically infected baboons produced higher levels of IgG and IgM than acute ones during the infection. Low production of these antibody isotypes could have been the reason why acutely infected baboons were not able to keep their parasitaemia levels low. There is growing evidence concerning the protective role of IgG in *P. falciparum* infections and passive transfer of IgG has provided protection against *P. falciparum* blood stage in South American monkeys (Groux and Gysin, 1990). Furthermore, human antibodies efficiently inhibit *in vitro P. falciparum* merozoite proliferation, and mediate opsonisation of infected erythrocytes (Groux and Gysin, 1990). There is evidence from human studies that antibodies have a role in protective immunity against malaria (Leoratti *et al.,* 2008). Results from the current study show that that increased levels of IgG and IgM have a protective role in immunity against *P. knowlesi* in baboons. The roles of cytokines in the regulation of immune responses against *Plasmodium* infections and pathogenesis of malaria has been extensively studied in rodent malaria models but there are few studies on cytokine studies in primates especially after experimental malaria (Yang *et al*., 1999). In the current study acutely infected baboons produced higher levels of IFN γ than chronically infected baboons. Production of high levels of IFN γ could have resulted in pathological changes that resulted in increased parasitaemia in acutely infected baboons. In one study in which human volunteers were infected experimentally with *Plasmodium falciparum,* there was increased production of IFN γ (Harpaz *et al.,* 1992). In a recent study (Adrande *et al.,* 2010), clinical presentations of *P. vivax* malaria infection were discovered to be strongly associated with a potent activation of pro-inflammatory responses and cytokine imbalance. Increased concentrations of the inflammatory markers IFN γ and TNF α were detected during severe *P. vivax* malaria infections in patients in the Brazilian Amazon. From our current study it is evident that high concentrations of IFN γ increase the severity *P. knowlesi* malaria infection in olive baboons (*Papio anubis*). This may not be a surprising outcome, considering that *P. knowlesi* is phylogenetically proximal to *P. vivax* (Ozwara *et al.,* 2003). Interferon gamma is an immune marker of pro-inflammatory responses that has previously been implicated in malaria disease immunopathology (Wroczynska *et al.,* 2005).

Acutely infected baboons produced higher level of IL 6 and TNF α than chronically infected baboons. These findings show that increased concentrations of these cytokines are markers of increased *P. knowlesi* malaria parasitisation and infection severity in the olive baboon model. Inflammation-mediated systemic damage may explain the severity of the baboons' clinical presentations. Interleukin 6 and TNF α were previously found to be markers of complicated *Plasmodium berghei* ANKA strain malaria (Grau *et al.,* 1990; Che *et al.,* 2000). High IL 6 concentrations were observed in mice with full-blown neurological syndrome during *P. berghei* ANKA strain malaria (Grau *et al.,* 1990). Excessive production of TNF-alpha may mediate the expression of ICAM-1 (Inter-Cellular Adhesion Molecule-1) on brain endothelial cells (EC) and hence cause the development of cerebral malaria (Che *et al.,* 2000). This outcome suggests that elevated concentrations of these cytokines in circulation elicited effective immune responses that were capable of maintenance of low levels of parasitaemia and infection severity.

In a previous study, plasma levels of IL-10, a cytokine that down-regulates inflammation, were detectably lower with increased malaria disease severity (Adrande *et al.,* 2010). Data obtained from cytokine administration and neutralization with antibodies shows that these cytokines are important in reduction of parasitaemia (Angulo and Fresno, 2002; Torre *et al.,* 2002). Results we have reported here show that these cytokines (IL 4, IL 10 and IL 12) have a protective role against *P. knowlesi* infection in baboons. It can be concluded from our study that acute *P. knowlesi* malaria is accompanied by increased concentrations of IFN γ, TNF α and IL 6 and reduced levels of circulating IL 10, IL 4, IL 12, IgG and IgM in the baboon host. Increased levels of IL 10, IL 4, IL 12, IgM and IgG and reduced IFN γ, TNF α and IL 6 levels are associated with increased protection against *P. knowlesi* malaria. These results are largely agreeable with data reported from human studies, thereby increasing the relevance of the olive baboon - *P. knowlesi* experimental infection system in future malaria studies.

## ACKNOWLEDGEMENT

This study was funded by the research capability strengthening World Health Organisation (WHO) grant (Grant Number: A 50075) for malaria research in Africa under the Multilateral Initiative on Malaria/Special Programme for Research and Training in Tropical Diseases (WHO-MIM/TDR). Special thanks to the entire Animal Resources Department (ARD) at the Institute of Primate Research (IPR) for providing and maintaining the olive baboons and other forms of support that they offered during the study.